\RequirePackage{ifpdf}
\ifpdf 
\documentclass[pdftex]{sigma}
\else
\documentclass{sigma}
\fi

\newcommand{\ud}{\, \mathrm{d}}
\newcommand{\Real}{\mathbb{R}}

\newcommand{\Natural}{\mathbb{N}}

\numberwithin{equation}{section}

\begin{document}

\allowdisplaybreaks
	
\renewcommand{\PaperNumber}{108}

\FirstPageHeading

\renewcommand{\thefootnote}{$\star$}

\ShortArticleName{Straight Quantum Waveguide with Robin Boundary
Conditions}

\ArticleName{Straight Quantum Waveguide \\ with Robin Boundary
Conditions\footnote{This paper is a
contribution to the Proceedings of the 3-rd Microconference
``Analytic and Algebraic Me\-thods~III''. The full collection is
available at
\href{http://www.emis.de/journals/SIGMA/Prague2007.html}{http://www.emis.de/journals/SIGMA/Prague2007.html}}}

\Author{Martin J\'ILEK}

\AuthorNameForHeading{M. J\'\i lek}

\Address{Faculty of Nuclear Sciences and Physical
Engineering, Czech Technical University,\\ B\v rehov\'a 7, 11519
Prague, Czech Republic}

\Email{\href{mailto:JilekM@km1.fjfi.cvut.cz}{JilekM@km1.fjfi.cvut.cz}}

\ArticleDates{Received August 10, 2007, in f\/inal form November 08, 2007; Published online November 21, 2007}

\Abstract{We investigate spectral properties of a quantum particle
conf\/ined to an inf\/inite straight planar strip by imposing Robin
boundary conditions with variable coupling. Assuming that the
coupling function tends to a constant at inf\/inity, we localize the
essential spectrum and derive a suf\/f\/icient condition which
guarantees the existence of bound states. Further properties of the
associated eigenvalues and eigenfunctions are studied numerically by
the mode-matching technique.}


\Classification{47F05; 47B25; 81Q05}

\section{Introduction}

Modern experimental techniques make it possible to fabricate tiny
semiconductor structures which are small enough to exhibit quantum
ef\/fects. These systems are sometimes called \emph{nano\-structures}
because of their typical size in a direction and they are expected
to become the building elements of the next-generation electronics.
Since the used materials are very pure and of crystallic structure,
the particle motion inside a nanostructure can be modeled by a free
particle with an ef\/fective mass $m^*$ living in a spatial region
$\Omega$. That is, the quantum Hamiltonian can be identif\/ied with
the operator
\begin{gather}
  H = -\frac{\hbar^2}{2m^*} \Delta \label{quantumhamiltonian}
\end{gather}
in the Hilbert space $L^2(\Omega)$, where $\hbar$ denotes the Planck
constant. We refer to \cite{DE,LCM} for more information on the
physical background.

An important category of nanostructures is represented by
\emph{quantum waveguides}, which are modeled by $\Omega$ being an
inf\/initely stretched tubular region in $\Real^2$ or $\Real^3$. In
principle, one can consider various conditions on the boundary of
$\Omega$ in order to model the fact that the particle is conf\/ined
to~$\Omega$. However, since the particle wavefunctions~$\psi$ are
observed to be suppressed near the interface between two dif\/ferent
semiconductor materials, one usually imposes Dirichlet boundary
conditions, i.e.\ $\psi = 0$ on $\partial\Omega$. Such models
were extensively studied. The simplest possible system is a straight
tube. The spectral properties of corresponding Hamiltonian in this
case are trivial in the sense that the discrete spectrum is empty.

It is known, that a deviation from the straight tube can give rise
to non-trivial spectral properties like existence of bound states,
by bending it~\cite{CDFK, DE, ES, GJ, KK, LJ, OM}, introducing an
arbitrarily small `bump'~\cite{BEGK, BGRS} or impurities modeled by
Dirac interaction~\cite{EK}, coupling several waveguides by a
window~\cite{ESTV}, etc.

Another possibility of generating bound states is the changing of
boundary conditions. It can be done by imposing a combination of
Dirichlet and Neumann boundary conditions on dif\/ferent parts of the
boundary. Such models were studied in~\cite{DKriz, DK, FK, KovKrej}.

In this paper, we introduce and study the model of straight planar
quantum waveguide with Robin conditions on the boundary. While to
impose the Dirichlet boundary conditions means to require the
vanishing of wavefunction on the boundary of $\Omega$, the Robin
conditions correspond to the weaker requirement of vanishing of the
probability current, in the sense that its normal component vanishes
on the boundary, i.e.
\[
  j \cdot n = 0 \qquad \textrm{on} \quad \partial\Omega,
\]
where the probability current $j$ is def\/ined by
\[
  j := \frac{i \hbar}{2 m^*} \left[ \psi \nabla \overline\psi -
  \overline\psi \nabla \psi \right].
\]
This less restrictive requirement may in principle model dif\/ferent
types of interface in materials, namely, a superconducting f\/ilm
sandwiched between two metallic non-superconducting substrates.

The Laplacian subject to Robin boundary conditions can be also used
in the problem to f\/ind the electro-magnetic f\/ield outside the object
consisting of a conducting core covered by a dielectric layer. If
the thickness of the layer is too small compared to the dimension of
the conductive core, the numerical methods of solving this problem
fail because of instabilities that then arise. In this case, the
problem can by solved by approximation of the dielectric layer by
appropriate boundary conditions of the Robin type. We refer to
\cite{BL,EN} for more information.

\begin{figure}[t]
 \centerline{\includegraphics[width=13cm]{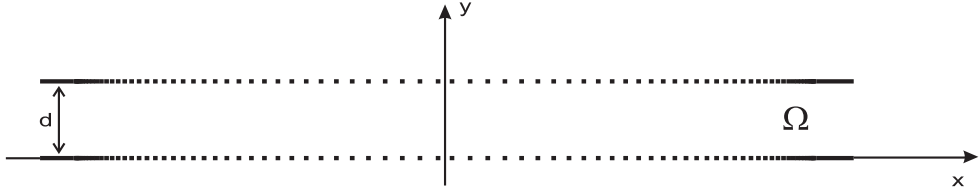}}
    \caption{Straight waveguide with Robin boundary conditions. The
    dotted lines indicate that the boundary-coupling function is
    allowed to vary.}
    \label{Fig.System}
\end{figure}

The system we are going to study is sketched on
Fig.~\ref{Fig.System}. We consider the quantum particle whose
motion is conf\/ined to a straight planar strip of width $d$. We shall
denote this conf\/iguration space by $\Omega := \Real \times (0, d)$.
For def\/initeness we assume that it is placed to the upper side of
the $x$-axis. On the boundary the Robin conditions are imposed. More
precisely, we suppose that every wave-function $\psi$ satisf\/ies
\begin{gather}
  -\partial_y \psi (x,0) + \alpha(x) \psi (x,0) = 0, \nonumber \\
  \partial_y \psi (x,d) + \alpha(x) \psi (x,d) = 0, \label{bc}
\end{gather}
for all $x \in \Real$. Notice that the parameter $\alpha$ depends on
the $x$-coordinate and this dependence is the same on both ``sides''
of the strip. We require that $\alpha(x)$ is positive for all $x \in
\Real$. Moreover, in Theorem~\ref{thm.sa} we will show that a
suf\/f\/icient condition for the self-adjointness of the Hamiltonian is
the requirement that $\alpha \in W^{1,\infty} (\Real)$.

Putting $\hbar^2 / 2 m^* = 1$ in \eqref{quantumhamiltonian}, we may
identify the particle Hamiltonian with the self-adjoint operator on
the Hilbert space $L^2(\Omega)$, def\/ined in the following way{\samepage
\begin{gather}
  H_\alpha \psi := -\Delta \psi, \qquad
  \forall\, \psi \in D(H_\alpha) :=
  \left\{ \psi \in W^{2,2} (\Omega) \,\big|\,
  \psi \ \mathrm{satisf\/ies} \ \eqref{bc}
  \right\}, \label{def.ham}
\end{gather}
where $D(H_\alpha)$ denotes the domain of the Hamiltonian.}

While it is easy to see that $H_\alpha$ is symmetric, it is quite
dif\/f\/icult to prove that it is self-adjoint. This will be done in the
next section. Section~\ref{sec.spectrum} is devoted to localization
of the essential spectrum and proving the existence of the discrete
spectrum. In the f\/inal section we study an example numerically to
illustrate the spectral properties.

\section{The self-adjointness of the Robin Laplacian}

For showing the self-adjointness of the Hamiltonian, we were
inspired by~\cite[Section~3]{BK}. Our strategy is to show that
$H_\alpha$ is, in fact, equal to another operator $\tilde H_\alpha$,
which is self-adjoint and def\/ined in following way.

Let us introduce a sesquilinear form
\begin{gather*}
  h_\alpha (\phi, \psi) := \int_\Omega \overline{ \nabla \phi (x,y)
  } \cdot \nabla \psi (x,y) \ud x \ud y + \int_\Real \alpha(x) \big(
  \overline{ \phi (x,0) } \psi (x,0) + \overline { \phi (x,d) } \psi
  (x,d) \big) \ud x
\end{gather*}
with the domain
\begin{gather*}
  D(h_\alpha) := W^{1,2} (\Omega).
\end{gather*}
Here the dot denotes the scalar product in $\Real^2$ and the
boundary terms should be understood in the sense of
traces~\cite[Section~4]{Adams}. We shall denote the corresponding
quadratic form by $h_\alpha [\psi] := h_\alpha (\psi, \psi)$. In
view of the f\/irst representation theorem~\cite[Theorem~VI.2.1]{Kato},
there exists the unique self-adjoint operator $\tilde H_\alpha$ in
$L^2(\Omega)$ such that $h_\alpha (\phi, \psi) = (\phi, \tilde
H_\alpha \psi)$ for all $\psi \in D(\tilde H_\alpha) \subset
D(h_\alpha)$ and $\phi \in D(h_\alpha)$, where
\begin{gather*}
  D(\tilde H_\alpha) = \left\{ \psi \in D(h_\alpha)\, \big|\, \exists\, F
  \in L^2(\Omega), \forall \,\phi \in D(h_\alpha), h_\alpha (\phi,
  \psi) = (\phi, F) \right\}.
\end{gather*}

For showing the equality between $H_\alpha$ and $\tilde H_\alpha$ we
will need the following result.

\begin{lemma}\label{lem.weak.sol}
Let $\alpha \in W^{1,\infty} (\Real)$ and $\forall\, x \in \Real$, $
\alpha(x) > 0$. For each $F \in L^2 (\Omega)$, a solution $\psi$ to
the problem
\begin{gather}
  \forall \, \phi \in W^{1,2} (\Omega), \qquad h_\alpha (\phi, \psi) = (\phi,
  F) \label{weak.sol}
\end{gather}
belongs to $D(H_\alpha)$.
\end{lemma}

\begin{proof}
For any function $\psi \in W^{1,2} (\Omega)$, we introduce the
dif\/ference quotient
\begin{gather*}
  \psi_\delta (x,y) := \frac { \psi (x+\delta,y) - \psi (x,y) }
  {\delta},
\end{gather*}
where $\delta$ is a small real number. Since
\begin{gather*}
  \left| \psi (x+\delta, y) - \psi (x,y) \right| = \left| \delta \int_0^1
  \partial_x \psi (x+\delta t, y) \ud t \right| \leq \left| \delta \right| \int_0^1
  \left| \partial_x \psi (x+ \delta t, y) \right| \ud t,
\end{gather*}
we get the estimate
\begin{gather*}
  \int_\Omega \left| \psi_\delta \right| ^2 \leq \int_\Omega \left(
  \int_0^1 \left| \partial_x \psi (x+\delta t, y) \right| \ud t \right) ^2
  \ud x \ud y \leq \int_\Omega \left( \int_0^1 \left| \partial_x \psi
  (x+\delta t, y) \right| ^2 \ud t \right) \ud x \ud y \\
\phantom{\int_\Omega \left| \psi_\delta \right| ^2} {} = \int_0^1 \left( \int_\Omega \left| \partial_x \psi (x+\delta
  t, y) \right| ^2 \ud x \ud y \right) \ud t =
  \int_\Omega \left| \partial_x \psi (x, y) \right| ^2 \ud x \ud y.
\end{gather*}
Therefore the inequality{\samepage
\begin{gather}
  \| \psi_\delta \|_{L^2(\Omega)} \leq \| \psi \|_{W^{1,2}(\Omega)}
  \label{est.def.ham}
\end{gather}
holds true.}

If $\psi$ satisf\/ies~\eqref{weak.sol}, then $\psi_\delta$ is a
solution to the problem
\begin{gather*}
  h_\alpha (\phi,\psi_\delta) = (\phi, F_\delta) -
  \int_\Real \alpha_\delta (x) \left( \overline{ \phi (x,0) } \psi (x+\delta,
  0) + \overline{ \phi (x,d) } \psi (x+\delta,
  d) \right) \ud x,
\end{gather*}
where $\phi \in W^{1,2} (\Omega)$ is arbitrary. Letting $\phi =
\psi_\delta$ and using the ``integration-by-parts'' formula for the
dif\/ference quotients, $(\phi, F_\delta) = - (\phi_{-\delta}, F)$, we
get
\begin{gather}
  h_\alpha [\psi_\delta] = -( (\psi_\delta)_{-\delta}, F) -
  \int_\Real \alpha_\delta (x) \left( \overline{ \psi_\delta (x,0) } \psi (x+\delta,
  0) + \overline{ \psi_\delta (x,d) } \psi (x+\delta,
  d) \right) \ud x. \label{eq.proof.sa}
\end{gather}
Using Schwarz inequality, Cauchy inequality, estimate
\eqref{est.def.ham}, boundedness of $\alpha$ and $\alpha_\delta$,
and embedding of $W^{1,2} (\Omega)$ in $L^2(\partial \Omega)$, we
can make following estimates
\begin{gather*}
  \left| ( (\psi_\delta)_{-\delta}, F) \right| \leq \| F \|_{L^2(\Omega)} \|
  (\psi_\delta)_{-\delta} \|_{L^2(\Omega)} \leq \frac{1}{2} \| F \|_{L^2(\Omega)}^2 +
  \frac{1}{2} \| \psi_\delta \|_{W^{1,2} (\Omega)}^2, \\
   \left| \int_\Real \alpha_\delta (x) \left(
  \overline{ \psi_\delta (x,0) } \psi (x+\delta,
  0) + \overline{ \psi_\delta (x,d) } \psi (x+\delta,
  d) \right) \ud x \right| \\
 \qquad {}  \leq
  C_1 \| \psi_\delta \|_{L^2(\partial\Omega)}
  \| \psi \|_{L^2(\partial\Omega)}   \leq C_2 \| \psi_\delta
  \|_{W^{1,2} (\Omega) } \| \psi \|_{W^{1,2} (\Omega) }, \\
  \left| \int_\Real \alpha(x) \left( |\psi_\delta (x,0)|^2 + |\psi_\delta
  (x,d)|^2 \right) \ud x \right| \leq
  C_3 \| \psi_\delta \|_{L^2(\partial\Omega)}^2 \leq C_4 \|
  \psi_\delta \|_{W^{1,2}(\Omega)}^2
\end{gather*}
with constants $C_1$--$C_4$ independent of $\delta$. Giving this
estimates together, the identity \eqref{eq.proof.sa} yields
\begin{gather*}
  -\| \psi_\delta\|_{W^{1,2}(\Omega)}^2 \leq
  C_4 \| \psi_\delta \|_{W^{1,2}(\Omega)}^2 +
  C_2 \| \psi_\delta \|_{W^{1,2} (\Omega) } \| \psi \|_{W^{1,2} (\Omega)
  } + \frac{1}{2} \| F \|_{L^2(\Omega)}^2 +
  \frac{1}{2} \| \psi_\delta \|_{W^{1,2} (\Omega)}^2.
\end{gather*}
We get the inequality
\begin{gather*}
  \| \psi_\delta \|_{W^{1,2} (\Omega)} \leq C,
\end{gather*}
where the constant $C$ is independent of $\delta$. This estimate
implies
\begin{gather*}
  \sup_\delta \| \psi_{-\delta} \|_{W^{1,2}(\Omega)} < \infty;
\end{gather*}
and, therefore, by~\cite[\S~D.4]{Evans} there exists a function $v
\in W^{1,2} (\Omega)$ and a subsequence $\delta_k \to 0$ such that
$\psi_{-\delta_k} \xrightarrow{w} v$ in $W^{1,2} (\Omega)$. But then
\begin{gather*}
  -\int_\Omega \partial_x \psi \phi = \int_\Omega \psi \partial_x \phi =
  \int_\Omega \psi \lim_{\delta_k \to 0} \phi_{\delta_k} =
  \lim_{\delta_k \to 0} \int_\Omega \psi \phi_{\delta_k} =
  -\lim_{\delta_k \to 0} \int_\Omega \psi_{-\delta} \phi = -
  \int_\Omega v \phi.
\end{gather*}
Thus, $\partial_x \psi = v$ in the weak sense, and so $\partial_x
\psi \in W^{1,2} (\Omega)$. Hence, $\partial_{xx} \psi \in
L^2(\Omega)$ and $\partial_{xy} \psi \in L^2(\Omega)$.

It follows from the standard elliptic regularity theorems
(see~\cite[\S~6.3]{Evans}) that $\psi \in W_{\rm loc}^{2,2} (\Omega)$.
Hence, the equation $-\Delta \psi = F$ holds true a.e.\ in $\Omega$.
Thus, $\partial_{yy} \psi = -F - \partial_{xx} \psi \in
L^2(\Omega)$, and therefore $\psi \in W^{2,2} (\Omega)$.

It remains to check boundary conditions for $\psi$. Using
integration by parts, one has
\begin{gather*}
  (\phi, F) = h_\alpha (\psi, \phi) = (\phi, -\Delta \psi)
 + \int_\Real \overline{\phi(x, 0)} \left[ -\partial_y \psi
  (x,0) + \alpha (x) \psi (x, 0) \right] \ud x \\
  \phantom{(\phi, F) = h_\alpha (\psi, \phi) =} {} + \int_\Real \overline{\phi(x, d)} \left[ \partial_y \psi
  (x,d) + \alpha (x) \psi (x, d) \right] \ud x
\end{gather*}
for any $\phi \in W^{1,2} (\Omega)$. This implies the boundary
conditions because $-\Delta \psi = F$ a.e.\ in $\Omega$ and $\phi$ is
arbitrary.
\end{proof}

\begin{theorem}\label{thm.sa}
Let $\alpha \in W^{1,\infty}(\Real)$ and $\forall\, x \in \Real, \
\alpha(x) > 0$. Then $\tilde H_\alpha = H_\alpha$.
\end{theorem}

\begin{proof}
Let $\psi \in D(H_\alpha)$, i.e., $\psi \in W^{2,2}(\Omega)$
and $\psi$ satisf\/ies the boundary conditions~\eqref{bc}. Then $\psi
\in D(h_\alpha) = W^{1,2}(\Omega)$ and by integration by parts
and~\eqref{bc} we get for all $\phi \in D(h_\alpha)$ the relation
\begin{gather*}
  h_\alpha(\phi,\psi) =
  \int_\Real \overline{\phi(x,d)} \partial_y \psi(x,d) \ud x -
  \int_\Real \overline{\phi(x,0)} \partial_y \psi(x,0) \ud x -
  \int_\Omega \overline{\phi(x,y)} \Delta \psi (x,y) \ud x \ud y \\
  \phantom{h_\alpha(\phi,\psi) =}{} +
  \int_\Real \alpha(x) \overline{\phi(x,d)}\psi(x,d) \ud x
 + \int_\Real \alpha(x) \overline{\phi(x,0)}\psi(x,0) \ud x\\
\phantom{h_\alpha(\phi,\psi)}{}  =
  -\int_\Omega \overline{\phi(x,y)} \Delta\psi(x,y) \ud x \ud y.
\end{gather*}
It means that there exists $\eta := -\Delta\psi \in L^2(\Omega)$
such that $\forall\, \phi \in D(h_\alpha)$, $h_\alpha(\phi, \psi) =
(\phi,\eta)$. That is, $\tilde H_\alpha$ is an extension of
$H_\alpha$.

The other inclusion holds as a direct consequence of
Lemma~\ref{lem.weak.sol} and the f\/irst representation theorem.
\end{proof}

\section{The spectrum of Hamiltonian}\label{sec.spectrum}

In this section we will investigate the spectrum of the Hamiltonian
with respect to the behavior of the function $\alpha$. In whole
section we suppose that $\alpha \in W^{1,\infty} (\Real)$ and
$\forall\, x \in \Real$, $\alpha(x) > 0$. We start with the simplest
case.

\subsection{Unperturbed system}

If $\alpha(x) = \alpha_0 > 0$ is a constant function, the
Schr\"odinger equation can be easily solved by separation of
variables. The spectrum of the Hamiltonian is then
$\sigma(H_{\alpha_0}) = [E_1(\alpha_0), \infty)$, where
$E_1(\alpha_0)$ is the f\/irst transversal eigenvalue. The transversal
eigenfunctions have the form
\begin{gather}
  \chi_n (y; \alpha) = N_\alpha \left( \frac {\alpha} {\sqrt{ E_n (\alpha) }}
  \sin \left( \sqrt{ E_n (\alpha) } y \right) +
  \cos \left( \sqrt{ E_n (\alpha) } y \right) \right),
  \label{eq.chi}
\end{gather}
where $N_\alpha$ is a normalization constant and the eigenvalues
$E_n(\alpha)$ are determined by the implicit equation
\begin{gather*}
  f (E_n; \alpha) = 2 \alpha \sqrt{E_n(\alpha)} \cos (\sqrt{E_n(\alpha)} d ) +
  (\alpha^2 - E_n(\alpha)) \sin (\sqrt{E_n(\alpha)} d ) = 0.
\end{gather*}
Note that there are no eigenvalues below the bottom of the essential
spectrum, i.e., $\sigma_{\rm disc} = \varnothing$.

\subsection{The stability of essential spectrum}

As we have seen, if $\alpha$ is a constant function, the essential
spectrum of the Hamiltonian is the interval $[E_1 (\alpha_0),
\infty)$. Now we prove that the same spectral result holds if
$\alpha$ tends to $\alpha_0$ at inf\/inity.

\begin{theorem} \label{thm.ess}
If $\lim\limits_{|x| \to \infty} \alpha(x) - \alpha_0 = 0$ then
$\sigma_{\rm ess} (H_\alpha) = [ E_1 (\alpha_0), \infty)$.
\end{theorem}

The proof of this theorem is achieved in two steps. Firstly, in
Lemma~\ref{lem.ess2}, we employ a Neumann bracketing argument to
show that the threshold of essential spectrum does not descent below
the energy $E_1(\alpha_0)$. Secondly, in Lemma~\ref{lem.ess1}, we
prove that all values above $E_1(\alpha_0)$ belongs to the essential
spectrum by means of the following characterization of the essential
spectrum which we have adopted from~\cite{DDI}.

\begin{lemma}\label{lem.char.ess}
Let $H$ be a non-negative self-adjoint operator in a complex Hilbert
space~$\mathcal{H}$ and~$h$ be the associated quadratic form. Then
$\lambda \in \sigma_{\rm ess} (H)$ if and only if there exists a
sequence $\{ \psi_n \}_{n=1}^\infty \subset D(h)$ such that
\begin{enumerate}\itemsep=0pt
\item[(i)]
$ \forall\, n \in \Natural \setminus \{ 0 \}$, $\| \psi_n \| =
1$,
\item[(ii)]
$\psi_n \xrightarrow[n \to \infty]{w} 0 \ \mathrm{in} \
\mathcal{H}$,
\item[(iii)]
$(H - \lambda) \psi_n \xrightarrow[n \to \infty]{ } 0 \
\mathrm{in} \ D(h)^*$.
\end{enumerate}
\end{lemma}
\noindent Here $D(h)^*$ denotes the dual of the space $D(h)$. We
note that
\begin{gather*}
  \| \psi \|_{D(h)^*} = \sup_{\phi \in D(h) \setminus \{0\}} \frac
  {|(\phi,\psi)|} {\| \phi \|_1 }
\end{gather*}
with
\begin{gather*}
  \| \phi \|_1 := \sqrt{ h[\phi] + \| \phi \|^2 }.
\end{gather*}

The main advantage of Lemma~\ref{lem.char.ess} is that it requires
to f\/ind a sequence from the form domain of $H$ only, and not from
$D(H)$ as it is required by the Weyl
criterion~\cite[Lemma~4.1.2]{Davies}. Moreover, in order to check the
limit from~{\it (iii)}, it is still suf\/f\/icient to consider the
operator $H$ in the form sense, i.e.\ we will not need to
assume that $\alpha$ is dif\/ferentiable in our case.

\begin{lemma}\label{lem.ess2}
If $\lim\limits_{|x| \to \infty} \alpha(x) - \alpha_0 = 0$ then $\inf
\sigma_{\rm ess} (H_\alpha) \geq E_1(\alpha_0)$.
\end{lemma}

\begin{proof}
Since $\alpha (x) - \alpha_0$ vanishes at inf\/inity, for any f\/ixed
$\varepsilon > 0$ there exists $a > 0$ such that
\begin{gather}
  | x | > a \ \Rightarrow \ | \alpha(x) - \alpha_0 | <
  \varepsilon. \label{limit.ess}
\end{gather}
Cutting $\Omega$ by additional Neumann boundary parallel to the
$y$-axis at $x = \pm a$, we get new operator $H_\alpha^{(N)}$
def\/ined using quadratic form. We can decompose this operator
$
  H_\alpha^{(N)} = H_{\alpha,t}^{(N)} \oplus H_{\alpha,c}^{(N)},
$
where the ``tail'' part corresponds to the two halfstrips ($|x|>a$)
and the rest to the central part with the Neumann condition on the
vertical boundary. Using Neumann bracketing, cf.~\cite[Section~XIII.15]{Reed}, we get
$
  H_\alpha^{(N)} \leq H_\alpha
$
in the sense of quadratic forms.

We denote
$
  \alpha_{\min} (a) := \inf\limits_{|x|>a} \alpha(x).
$
Since $\sigma_{\rm ess} (H_{\alpha_{\min},t}^{(N)}) = [E_1(\alpha_{\min}),
\infty)$ and
$
  H_{\alpha_{\min},t}^{(N)} \leq H_{\alpha,t}^{(N)}
$
in the sense of quadratic forms, we get the following estimate of
the bottom of the essential spectrum of the ``tail'' part
\begin{gather}
  E_1 (\alpha_{\min}) \leq \inf \sigma_{\rm ess} (H_{\alpha,t}^{(N)}).
  \label{opineq.ess}
\end{gather}
Since the spectrum of $H_{\alpha,c}^{(N)}$ is purely discrete, cf.~\cite[Chapter~7]{Davies}, the minimax principle gives the inequality
\begin{gather}
  \inf \sigma_{\rm ess} (H_{\alpha,t}^{(N)}) \leq \inf \sigma_{\rm ess}
  (H_\alpha). \label{infineq.ess}
\end{gather}
Since the assertion~\eqref{limit.ess} yields
$
  \alpha_0 - \varepsilon < \alpha_{\min}
$
and $E_1$ is an increasing function of $\alpha$, we have
\begin{gather}
  E_1(\alpha_0 - \varepsilon) < E_1(\alpha_{\min}). \label{e1.ess}
\end{gather}
Giving together~\eqref{opineq.ess}, \eqref{infineq.ess}, and
\eqref{e1.ess} we get the relation
$
  E_1(\alpha_0 - \varepsilon) < \inf \sigma_{\rm ess} (H_\alpha).
$
The claim then follows from the fact that  $E_1 = E_1 (\alpha)$ is a
continuous function.
\end{proof}

\begin{lemma}\label{lem.ess1}
If $\lim\limits_{|x| \to \infty} \alpha(x) - \alpha_0 = 0$ then $ [ E_1
(\alpha_0), \infty) \subseteq \sigma_{\rm ess} (H_\alpha)$.
\end{lemma}

\begin{proof}
Let $\lambda \in [E_1 (\alpha_0), \infty)$. We shall construct a
sequence $\left\{ \psi_n \right\}_{n=1}^\infty$ satisfying the
assumptions~{\it (i)}--{\it (iii)} of Lemma~\ref{lem.char.ess}. We
def\/ine the following family of functions
\begin{gather*}
  \psi_n (x,y) := \varphi_n (x) \chi_1 (y ; \alpha_0) \exp \left( i
  \sqrt{\lambda - E_1 (\alpha_0)} x \right),
\end{gather*}
where $\chi_1$ is the lowest transversal function~\eqref{eq.chi} and
$\varphi_n (x) := n^{-1/2} \varphi (x/n - n)$ with $\varphi$
satisfying
\begin{enumerate}\itemsep=0pt
\item[1)] $\varphi \in C_0^\infty (\Real)$,
\item[2)] $\forall\,  x \in \Real$, $0 \leq \varphi (x) \leq 1$,
\item[3)] $\forall \, x \in \left( -1/4, 1/4 \right)$, $\varphi = 1$,
\item[4)] $\forall \,x \in \Real \setminus \left[ -1/4, 1/4 \right]$, $\varphi = 0$,
\item[5)] $ \| \varphi \|_{L^2(\Real)} = 1$.
\end{enumerate}
Note that $\mathrm{supp} \ \varphi_n \subset (n^2 - n, n^2 + n)$. It
is clear that $\psi_n$ belongs to the form domain of $H_\alpha$. The
assumption~{\it (i)}  of Lemma~\ref{lem.char.ess} is satisf\/ied due
to the normalization of~$\chi_1$ and~$\varphi$.

The point~{\it (ii)} of Lemma~\ref{lem.char.ess} requires that
$(\phi, \psi_n) \to 0$ as $n \to \infty$ for all $\phi \in
C_0^\infty (\Omega)$, a dense subset of $L^2(\Omega)$. However, it
follows at once because $\phi$ and $\psi_n$ have disjoint supports
for $n$ large enough.

Hence, it remains to check that $\| (H_\alpha - \lambda) \psi_n
\|_{D(h)^*} \to 0$ as $n \to \infty$. An explicit calculation using
integration by parts, boundary conditions \eqref{bc} and the
relations
\begin{gather}
\| \dot \varphi_n \|_{L^2(\Real)} = n^{-1} \| \dot \varphi
\|_{L^2(\Real)}, \qquad \| \ddot \varphi_n \|_{L^2(\Real)} = n^{-2}
\| \ddot \varphi \|_{L^2(\Real)} \label{eq.norm.varphi}
\end{gather}
yields
\begin{gather*}
  \big| \big( \phi, (H_\alpha - \lambda) \psi_n \big) \big| =
  \bigg| \int_\Omega \overline{\phi(x,y)} \chi_1(y;\alpha_0) \ddot
  \varphi_n (x) \exp \left( i \sqrt{\lambda - E_1 (\alpha_0)} x
  \right) \ud x \ud y \\
\qquad{} + 2 i \int_\Omega \overline{\phi(x,y)}
  \chi_1(y;\alpha_0) \dot \varphi_n (x) \sqrt{\lambda - E_1
  (\alpha_0)} \exp \left( i \sqrt{\lambda - E_1 (\alpha_0)} x
  \right) \ud x \ud y \bigg| \\
  \qquad {} \leq n^{-2} \| \phi \|_{L^2(\Omega)}
  \| \chi_1 \|_{L^2((0,d))} \| \ddot \varphi \|_{L^2(\Real)}
   + 2 n^{-1} \| \phi \sqrt{\lambda {-} E_1 (\alpha_0)}
  \|_{L^2(\Omega)} \| \chi_1 \|_{L^2((0,d))} \| \dot \varphi \|_{L^2(\Real)}
\end{gather*}
for all $\phi \in D(h_\alpha)$. The claim then follows from the fact
that both terms at the r.h.s.\ go to zero as $n \to \infty$.
\end{proof}

\subsection{The existence of bound states}

Now we show that if $\alpha - \alpha_0$ vanishes at inf\/inity, some
behavior of the function $\alpha$ may produce a non-trivial spectrum
below the energy $E_1(\alpha_0)$. Note that this together with the
assumption of Theorem~\ref{thm.ess} implies that the spectrum below
$E_1 (\alpha_0)$ consists of isolated eigenvalues of f\/inite
multiplicity, i.e.\ $\sigma_{\rm disc} (H_\alpha) \neq \varnothing$.
Suf\/f\/icient condition that pushes the spectrum of the Hamiltonian
below $E_1 (\alpha_0)$ is introduced in the following theorem.

\begin{theorem}\label{thm.bs}
Suppose
\begin{enumerate}
\itemsep=0pt
\item[1)] $\alpha(x) - \alpha_0 \in L^1(\Real)$,
\item[2)] $\int_\Real \left( \alpha(x) - \alpha_0 \right) \ud x < 0$.
\end{enumerate}
Then $\inf \sigma(H_\alpha) < E_1(\alpha_0)$.
\end{theorem}

\begin{proof}
The proof is based on the variational strategy of f\/inding a trial
function $\psi$ from the form domain of $H_\alpha$ such that
\begin{gather}
  Q_\alpha [\psi] :=
  h_\alpha [\psi] - E_1(\alpha_0) \| \psi \|_{L^2(\Omega)}^2 < 0. \label{ineq.bs}
\end{gather}
Inspired by~\cite{GJ}, we def\/ine a sequence
\begin{gather*}
  \psi_n (x,y) := \varphi_n (x) \chi_1(y; \alpha_0),
  \qquad
  \varphi_n (x) := n^{-1/2} \varphi (x/n),
\end{gather*}
where the function $\varphi$ was def\/ined in the proof of
Lemma~\ref{lem.ess1}. Using the relations \eqref{eq.norm.varphi} and
the integration by parts we get
\begin{gather*}
  Q_\alpha [\psi_n] = n^{-2} \| \dot \varphi \|_{L^2(\Real)}^2 + \left (
  | \chi_1 (0; \alpha_0) |^2 + | \chi_1 (d; \alpha_0) |^2 \right)
  \int_\Real (\alpha (x) - \alpha_0) \varphi_n (x) \ud x.
\end{gather*}
Since the integrand is dominated by the $L^1$-norm of
$\alpha-\alpha_0$ we have the limit
\begin{gather*}
  \lim_{n \to \infty} Q_\alpha [\psi_n] =\left (
  | \chi_1 (0; \alpha_0) |^2 + | \chi_1 (d; \alpha_0) |^2 \right)
  \int_\Real (\alpha (x) - \alpha_0) \ud x
\end{gather*}
by dominated convergence theorem. This expression is negative
according to the assumptions. Now, it is enough to take $n$
suf\/f\/iciently large to satisfy inequality~\eqref{ineq.bs}.
\end{proof}

\section{A `rectangular well' example}

To illustrate the above results and to understand the behavior of
the spectrum of the Hamiltonian in more detail, we shall now
investigate an example. Inspired by~\cite{EK} we choose the function
$\alpha$ to be a steplike function which makes it possible to solve
the corresponding Schr\"odinger equation numerically by employing
the mode-matching method. The simplest non-trivial case concerns a
`rectangular well' of a width $2 a$,
\begin{gather*}
  \alpha (x) = \left\{
  \begin{array}{ll}
    \alpha_1  & \mathrm{if} \ |x| < a,\vspace{1mm} \\
    \alpha_0       & \mathrm{if} \ |x| \geq a
  \end{array}
  \right.
\end{gather*}
with $a>0$ and $0 < \alpha_1 < \alpha_0$. In view of Theorem
\ref{thm.bs} this waveguide system has bound states. In particular,
one expects that in the case when $\alpha_1$ is close to zero and
$\alpha_0$ is large the spectral properties will be similar to those
of the situation studied in~\cite{DKriz}.

Since the system is symmetric with respect to the $y$-axis, we can
restrict ourselves to the part of $\Omega$ in the f\/irst quadrant and
we may consider separately the symmetric and antisymmetric
solutions, i.e.\ to analyze the halfstrip with the Neumann or
Dirichlet boundary condition at the segment $(0, d)$ on $y$-axis,
respectively.

\subsection{Preliminaries}
Theorem \ref{thm.ess} enables us to localize the essential spectrum
in the present situation, i.e., $\sigma_{\rm ess}(H_\alpha) =
[E_1(\alpha_0), \infty)$. Moreover, according to the minimax
principle we know that isolated eigenvalues of $H_\alpha$ are
squeezed between those of $H_{a,c}^{(N)}$ and $H_{a,c}^{(D)}$, the
Hamiltonians in the central part with Neumann or Dirichlet condition
on the vertical boundary, respectively. The Neumann estimate tells
us that $\inf \sigma(H_\alpha) \geq E_1(\alpha_1)$. One f\/inds that
the $n$-th eigenvalue $E_n$ of $H_\alpha$ is estimated by
\begin{gather*}
  E_1(\alpha_1) + \left( \frac{(n-1) \pi}{2a} \right)^2 \leq
  E_n \leq E_1(\alpha_1) + \left( \frac{n \pi}{2a} \right)^2.
\end{gather*}

\subsection{Mode-matching method}

Let us pass to the mode-matching method. A natural Ansatz for the
solution of an energy $\lambda \in ( E_1 (\alpha_1), E_1 (\alpha_0)
)$ is
\begin{gather*}
  \psi_{s/a} (x,y) = \sum_{n=1}^\infty a_n^{s/a}
  \left\{ \begin{array}{c}
    \frac {\cosh(l_n x)} {\cosh(l_n a)} \vspace{2mm}\\
    \frac {\sinh(l_n x)} {\sinh(l_n a)}
  \end{array} \right\} \chi_n (y; \alpha_1)
  \qquad \mathrm{for} \ 0 \leq x < a, \\
  \psi_{s/a} (x,y) = \sum_{n=1}^\infty b_n^{s/a}
  \exp(-k_n(x-a)) \chi_n (y; \alpha_0)
  \qquad \mathrm{for} \ x \geq a,
\end{gather*}
where the subscripts and superscripts $s$, $a$ distinguish the
symmetric and antisymmetric case, respectively. The longitudinal
momenta are def\/ined by
\begin{gather*}
  l_n := \sqrt{E_n(\alpha_1) - \lambda}, \qquad
  k_n := \sqrt{E_n(\alpha_0) - \lambda}.
\end{gather*}

\begin{figure}[t]
\centerline{\includegraphics[width = 13cm]{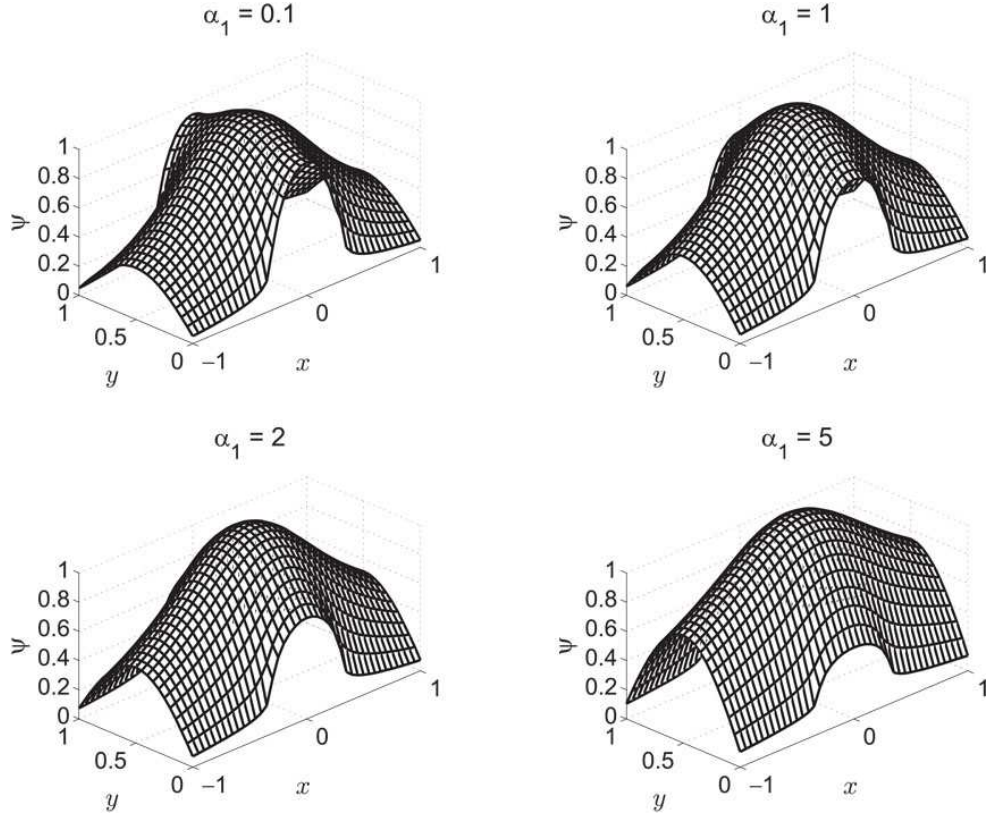}}
    \caption{Ground state eigenfunctions for $a/d = 0.3$, $\alpha_0 = 20$.}
    \label{fig.eigenfunctions}
\end{figure}

As an element of the domain \eqref{def.ham}, the function $\psi$
should be continuous together with its normal derivative at the
segment dividing the two regions, $x = a$. Using the orthonormality
of~$\{ \chi_n \}$ we get from the requirement of continuity
\begin{gather}
  \sum_{n=1}^\infty a_n \int_0^d \chi_n(y,\alpha_1)
  \chi_m(y;\alpha_0) \ud y = b_m. \label{eq.cont}
\end{gather}
In the same way, the normal-derivative continuity at $x = a$ yields
\begin{gather}
  \sum_{n=1}^\infty a_n l_n
  \left\{ \begin{array}{c}
    \tanh \\ \coth
  \end{array} \right\}
  (l_n a) \int_0^d \chi_n(y;\alpha_1)
  \chi_m(y; \alpha_0) \ud y + b_m k_m = 0. \label{eq.deriv}
\end{gather}
Substituting \eqref{eq.cont} to \eqref{eq.deriv} we can write the
equation as
\begin{gather}
  \mathbf{C a} = \mathbf{0}, \label{eq.mm}
\end{gather}
where
\begin{gather*}
  C_{mn} = \left( l_n
  \left\{ \begin{array}{c}
    \tanh \\ \coth
  \end{array} \right\}
  (l_n a) + k_m \right)
  \int_0^d \chi_n(y;\alpha_1)
  \chi_m(y; \alpha_0) \ud y.
\end{gather*}

In this way we have transformed a partial-dif\/ferential-equation
problem to a solution of an inf\/inite system of linear equations. The
latter will be solved numerically by using in~\eqref{eq.mm} an $N$
by $N$ subblock of $\mathbf{C}$ with large $N$.

\begin{figure}[t]
\centerline{\includegraphics[width = 13cm]{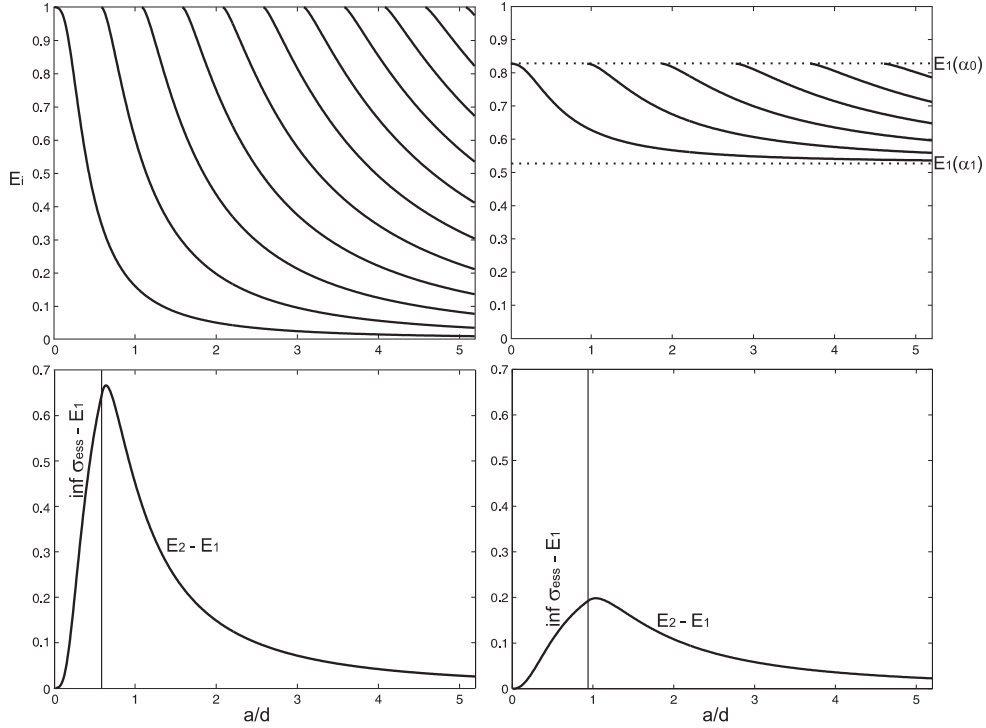}}
    \caption{The dependence of eigenvalues (in the units of $(\pi/d)^2$) on
    $a/d$ in the case $\alpha_0 =
    10^5$, $\alpha_1 = 10^{-5}$ (left) and $\alpha_0 = 20$,
    $\alpha_1 = 5$ (right) and corresponding f\/irst gaps.}
    \label{fig.eigenvalues}
\end{figure}

\begin{figure}[th]
\centerline{\includegraphics[width = 10cm]{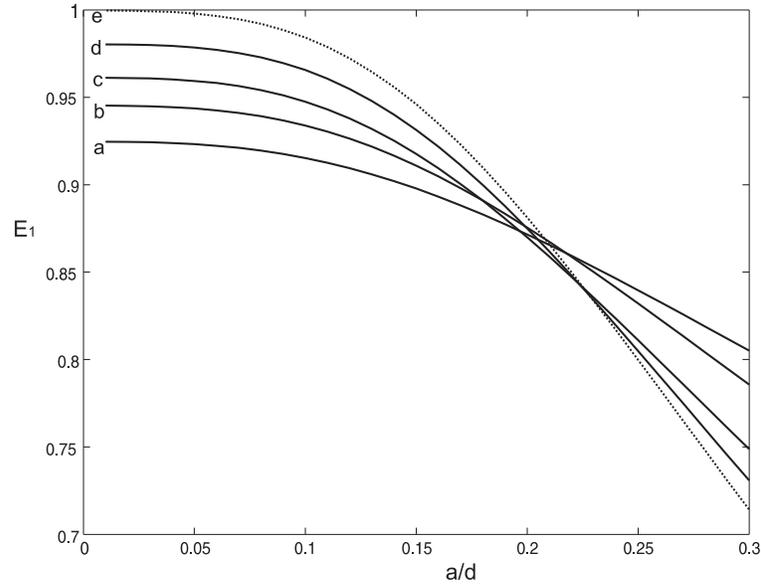}}
    \caption{The dependence of f\/irst eigenvalue (in the units of $(\pi/d)^2$) on
    $a/d$ in the cases: (a) $\alpha_0 = 50, \alpha_1=3$, (b) $\alpha_0 = 70, \alpha_1=2$,
    (c) $\alpha_0 = 100, \alpha_1=1$, (d) $\alpha_0 = 200, \alpha_1=0.5$, and in the
    Dirichlet--Neumann case (e) adopted from~\cite{DKriz}.}
    \label{fig.firsteigenvalue}
\end{figure}

\subsection{Numerical results}
The dependence of the ground state wavefunction with respect to
$\alpha_1$ for the central part of the halfwidth $a/d=0.3$ is
illustrated in Fig.~\ref{fig.eigenfunctions}. The slope of the
wavefunctions at the boundary between dif\/ferent $\alpha_i$ becomes
smoother with increasing $\alpha_1$, at $\alpha_1$ approaching
$\alpha_0$ the wavefunction is spread further and further in the
$x$-direction, and at $\alpha_1=\alpha_0$ bound state disappears.

Fig.~\ref{fig.eigenvalues} shows the bound-state energies as
functions of the `window' halfwidth $a/d$ in the case $\alpha_0 =
10^5$, $\alpha_1 = 10^{-5}$ (on the left) and $\alpha_0 = 20$,
$\alpha_1 = 5$ (on the right). We see that the energies (in both
cases) decrease monotonously with the increasing `window' width and
in the limit approach the uniform value $E_1(\alpha_1)$. Their
number increases as a function of $a/d$. Comparing this f\/igure with
Fig.~4 in~\cite{LJ} or Fig.~5 in~\cite{OM}, where the authors study
the planar waveguide comprised of two straight regions connected by
a region of constant curvature (with Dirichlet or mixed boundary
conditions, respectively), we may see the similar behavior of the
eigenvalues below the fundamental propagation constant. This
comparison shows a community of the physical processes taking place
in dif\/ferent structures where the longitudinal uniformity of the
waveguide is broken by some obstacle~-- either by the dif\/ferent
boundary conditions or the bend. For illustration, the f\/irst gap,
i.e., the dif\/ference between f\/irst and second eigenvalue (or
between f\/irst eigenvalue and the bottom of the essential spectrum if
there is only one eigenvalue), is plotted in the bottom part of
Fig.~\ref{fig.eigenvalues}. It is interesting that the dependence
on $a/d$ in the region where more than one eigenvalue exists is not
monotonous, but there is a maximum.

We can compare the dependence of the f\/irst eigenvalue on $a/d$ with
the system studied in~\cite{DKriz}. The authors there are interested
in the model of straight quantum waveguide with combined Dirichlet
and Neumann boundary conditions. Fig.~\ref{fig.firsteigenvalue}
shows that the larger $\alpha_0$ and the smaller~$\alpha_1$ is
(curves $a$, $b$, $c$, and $d$), the closer to the Dirichlet--Neumann
case (curve $e$) is the behavior of the f\/irst eigenvalue as a
function of $a/d$.

\subsection*{Acknowledgements}

The author thanks the referees for helpful suggestions.

\pdfbookmark[1]{References}{ref}
\LastPageEnding

\begin{thebibliography}{99}

\footnotesize\itemsep=0pt

\bibitem{Adams}
Adams R.A., Sobolev spaces, Academic Press, New York, 1975.

\bibitem{BL}
Bendali A., Lemrabet K., The ef\/fect of a thin coating on the
scattering of a time-harmonic wave for the Helmholtz equation, {\it
SIAM J. Appl. Math.} {\bf 56} (1996), 1664--1693.

\bibitem{BEGK}
Borisov D., Exner P., Gadyl'shin R., Krej\v{c}i\v{r}\'\i k D.,
Bound states in weakly deformed strips and layers, {\it Ann. Henri
Poincar\'e} {\bf 2} (2001), 553--572, \href{http://arxiv.org/abs/math-ph/0011052}{math-ph/0011052}.

\bibitem{BK}
Borisov D., Krej\v ci\v r\'ik D., PT-symmetric waveguide,
\href{http://arxiv.org/abs/0707.3039}{arXiv:0707.3039}.

\bibitem{BGRS}
Bulla W., Gesztesy F., Renger W., Simon B., Weakly coupled
boundstates in quantum waveguides, {\it Proc. Amer. Math. Soc.} {\bf
127} (1997), 1487--1495.

\bibitem{CDFK}
Chenaud B., Duclos P., Freitas P., Krej\v{c}i\v{r}\'\i k D.,
Geometrically induced discrete spectrum in curved tubes, {\it
Differential Geom. Appl.} {\bf 23} (2005), 95--105, \href{http://arxiv.org/abs/math.SP/0412132}{math.SP/0412132}.

\bibitem{Davies}
Davies E.B., Spectral theory and dif\/ferential operators, Camb. Univ.
Press, Cambridge, 1995.

\bibitem{DDI}
Dermenjian Y., Durand M., Iftimie V., Spectral analysis of an
acoustic multistratif\/ied perturbed cylinder, {\it Comm. Partial
Differential Equations} {\bf 23} (1998), 141--169.

\bibitem{DKriz}
Dittrich J., K\v r\'i\v z J., Bound states in straight quantum
waveguides with combined boundary conditions, {\it J.~Math. Phys.}
{\bf 43} (2002) 3892--3915, \href{http://arxiv.org/abs/math-ph/0112018}{math-ph/0112018}.

\bibitem{DK}
Dittrich J., K\v{r}\'\i\v{z} J., Curved planar quantum wires with
Dirichlet and Neumann boundary conditions, {\it J.~Phys.~A: Math. Gen.} {\bf 35}
(2002), L269--L275, \href{http://arxiv.org/abs/math-ph/0203007}{math-ph/0203007}.

\bibitem{DE}
Duclos P., Exner P., Curvature-induced bound states in quantum
waveguides in two and three dimensions, {\it Rev. Math. Phys.} {\bf
7} (1995), 73--102.

\bibitem{EN}
Engquist B., Nedelec J.C., Ef\/fective boundary conditions for
electro-magnetic scattering in thin layers, {\it Rapport Interne},
Vol.~278, CMAP, 1993.

\bibitem{Evans}
Evans L.C., Partial dif\/ferential equations, American Mathematical
Society, Providence, 1998.

\bibitem{EK}
Exner P., Krej\v ci\v r\'ik D., Quantum waveguides with a lateral
semitransparent barrier: spectral and scattering properties, {\it J.
Phys. A: Math. Gen.} {\bf 32} (1999), 4475--4494, \href{http://arxiv.org/abs/cond-mat/9904379}{cond-mat/9904379}.

\bibitem{ES}
Exner P., \v{S}eba P., Bound states in curved quantum waveguides,
{\it J. Math. Phys.} {\bf 30} (1989), 2574--2580.

\bibitem{ESTV}
Exner P., \v Seba P., Tater M., Van\v ek D., Bound states and
scattering in quantum waveguides coupled laterally through a
boundary window, {\it J. Math. Phys.} {\bf 37} (1996), 4867--4887, \href{http://arxiv.org/abs/cond-mat/9512088}{cond-mat/9512088}.

\bibitem{FK}
Freitas P., Krej\v{c}i\v{r}\'\i k D., Waveguides with combined
Dirichlet and Robin boundary conditions, {\it Math. Phys. Anal.
Geom.} {\bf 9} (2006), 335--352, \href{http://arxiv.org/abs/math-ph/0701075}{math-ph/0701075}.

\bibitem{GJ}
Goldstone J., Jaf\/fe R.L., Bound states in twisting tubes, {\it
Phys. Rev.~B} {\bf 45} (1992), 14100--14107.

\bibitem{Kato}
Kato T., Perturbation theory for linear operators, Springer-Verlag,
Berlin, 1966.

\bibitem{KovKrej}
Kova\v{r}\'\i k H.,  Krej\v{c}i\v{r}\'\i k D., A Hardy inequality in a
twisted Dirichlet--Neumann waveguide, {\it Math. Nachr.}, to appear,
\href{http://arxiv.org/abs/math-ph/0603076}{math-ph/0603076}.

\bibitem{KK}
Krej\v{c}i\v{r}\'\i k D., K\v{r}\'\i\v{z} J., On the spectrum of curved planar waveguides, {\it Publ. RIMS Kyoto Univ.} {\bf 41}
(2005), 757--791.

\bibitem{LJ} Lin K., Jaf\/fe R.L., Bound states and quantum
resonances in quantum wires with circular bends, {\it Phys. Rev.~B}
{\bf 54} (1996), 5750--5762.

\bibitem{LCM} Londergan J.T., Carini J.P., Murdock D.P., Binding
and scattering in two-dimensional systems, {\it Lect. Note in Phys.},
Vol.~60, Springer, Berlin, 1999.

\bibitem{OM} Olendski O., Mikhailovska L., Localized-mode evolution in a curved planar waveguide with combined Dirichlet and
Neumann boundary conditions, {\it Phys. Rev. E} {\bf 67} (2003),
056625, 11~pages.

\bibitem{Reed}
Reed M., Simon B., Methods of modern mathematical physics.
IV.~Analysis of operators, Academic Press, New York, 1978.

\end{thebibliography}
\end{document}